\documentclass[letterpaper,11pt]{article}

\usepackage[english]{babel}
\usepackage{theorem}
\theoremheaderfont{\itshape\bfseries}
{\theorembodyfont{\itshape}

	\newtheorem{definition}{\textbf{Definition}}
	\newtheorem{theorem}{\textbf{Theorem}}
	
	\newtheorem{remark}{\textbf{Remark}}
	
	\newtheorem{problem}{\textbf{Problem}}
}

\usepackage{cite}
\usepackage{newtxtext,newtxmath}
\usepackage{graphicx}
\usepackage[most]{tcolorbox}

\newcommand{\R}{\mathbb{R}}
\newcommand{\C}{\mathbb{C}}

\let\leq\leqslant
\let\geq\geqslant

\newenvironment{proof}[1][Proof]%
{\par\noindent\textit{#1:\ }}%
{\hspace*{\fill} \rule{6pt}{6pt}}
\newenvironment{proof*}[1][Proof]%
{\par\noindent\textit{#1:\ }}{}

\DeclareMathOperator{\diag}{diag}

\DeclareMathOperator{\re}{Re}

\newenvironment{system}[1]%
{\setlength{\arraycolsep}{0.5mm}\left\{ \; \begin{array}{#1}}%
	{\end{array} \right.}
\newenvironment{system*}[1]%
{\setlength{\arraycolsep}{0.5mm} \begin{array}{#1}}%
	{\end{array}}

\begin{document}

\title{Fault-tolerant properties of scale-free linear protocols for synchronization of homogeneous multi-agent systems }

\author{Anton A. Stoorvogel, Ali Saberi, and Zhenwei Liu
	\thanks{Anton A. Stoorvogel is with Department of Electrical
		Engineering, Mathematics and Computer Science, University of
		Twente, P.O. Box 217, Enschede, The Netherlands (e-mail:
		A.A.Stoorvogel@utwente.nl)}
	\thanks{Ali Saberi is with
		School of Electrical Engineering and Computer Science, Washington
		State University, Pullman, WA 99164, USA (e-mail: saberi@wsu.edu)}
	\thanks{Zhenwei Liu is with College of Information Science and
		Engineering, Northeastern University, Shenyang 110819,
		China (e-mail: liuzhenwei@ise.neu.edu.cn)} 
} 

\maketitle

\begin{abstract}
  Originally, protocols were designed for multi-agent systems (MAS)
  using information about the network which might not be available.
  Recently, there has been a focus on scale-free synchronization where
  the protocol is designed without any prior information about the
  network.
  
  As long as the network contains a directed spanning tree, a
  scale-free protocol guarantees that the network achieves
  synchronization.
  
  If there is no directed spanning tree then synchronization cannot be
  achieved. But what happens when these scale-free protocols are
  applied to such a network where the directed spanning tree no longer
  exists? This paper establishes that the network decomposes into a
  number of basic bicomponents which achieves synchronization among
  all nodes in this basic bicomponent. On the other hand, nodes which
  are not part of any basic bicomponent converge to a weighted average
  of the synchronized trajectories of the basic bicomponents. The
  weights are independent of the initial conditions and are
  independent of the designed protocol.
\end{abstract}

\section{Introduction}

The synchronization problem for multi-agent systems (MAS) has
attracted substantial attention due to its potential for applications
in several areas, see for instance the books
\cite{bai-arcak-wen,bullobook,kocarev-book,mesbahi-egerstedt,ren-book,%
	saberi-stoorvogel-zhang-sannuti,wu-book} or the papers
\cite{li-duan-chen-huang,saber-murray2,ren-beard}.

Most of the proposed protocols in the literature for synchronization
of MAS require some knowledge of the communication network such as
bounds on the spectrum of the associated Laplacian matrix or the
number of agents. As it is pointed out in
\cite{studli-2017,tegling-bamieh-sandberg,%
	tegling-bamieh-sandberg-auto-23,tegling2019scalability}, these
protocols suffer from \textbf{scale fragility} where stability
properties are lost when the size of the network increases or when the
network changes due to addition or removal of links.

In the past few years, \textbf{scale-free} linear protocol design has
been actively studied in the MAS literature to deal with the
existing scale fragility in MAS
\cite{liu-nojavanzedah-saberi-2022-book}.  Scale-free design
implies that the protocols are  designed solely based on the
knowledge of agent models and do \textbf{not} depend on 
\begin{itemize}
	\item information about the communication network such as the spectrum
	of the associated Laplacian matrix or
	\item knowledge of the number of agents.
\end{itemize}
In, for instance, \cite{liu-nojavanzedah-saberi-2022-book}, scale-free
protocols have been designed utilizing localized information exchange
between agent and its neighbors. Due
to this local information exchange the protocols are referred to as
\textbf{collaborative} protocols.

By contrast, scale-free \textbf{non-collaborative} protocol designs
only use relative measurements and no additional information exchange
is available and the protocols are, therefore, fully distributed.  The
necessary and sufficient conditions for solvability of scale-free
design via non-collaborative linear protocol for MAS consisting of
SISO agent model were recently reported in
\cite{liu-saberi-stoorvogel-tac-2023}.

For both collaborative and non-collaborative protocol, the designs
achieve synchronization for any communication network which contains a
directed spanning tree. If, for instance, due to a faulty link, the
network no longer contains a directed spanning tree then
synchronization is no longer achieved. This in itself is not
surprising since the existence of a directed spanning tree is a
necessary condition for achieving network synchronization.

However, it is an interesting question what happens if we apply our
scale-free protocol to a network which no longer contains a directed
spanning tree. Can this cause instability where synchronization errors
blow up or is there some inherent stability in the system. This
question is answered in this paper for both collaborative and
non-collaborative scale-free protocols. We will establish that the
network can be decomposed in basic bicomponents and a set of
additional nodes.
\begin{itemize}
	\item Within each basic bicomponent we achieve synchronization.
	\item The additional nodes converge to a weighted average of the
	synchronized trajectories of the basic bicomponents. The nonnegative
	weights are independent of the initial conditions, sum up to $1$ and
	are independent of the designed protocol.
\end{itemize}
It will be established that this behavior is true for both
collaborative and non-collaborative protocols \textbf{independent} of their
specific design methodology. 

\section{Communication network and graph}

To describe the information flow among the agents we associate a
{weighted graph} $\mathcal{G}$ to the communication network. The
weighted graph $\mathcal{G}$ is defined by a triple
$(\mathcal{V}, \mathcal{E}, \mathcal{A})$ where
$\mathcal{V}=\{1,\ldots, N\}$ is a node set, $\mathcal{E}$ is a set of
pairs of nodes indicating connections among nodes, and
$\mathcal{A}=[a_{ij}]\in \mathbb{R}^{N\times N}$ is the weighted
adjacency matrix with non negative elements $a_{ij}$. Each pair in
$\mathcal{E}$ is called an edge, where $a_{ij}>0$ denotes an
edge $(j,i)\in \mathcal{E}$ from node $j$ to node $i$ with weight
$a_{ij}$. Moreover, $a_{ij}=0$ if there is no edge from node $j$ to
node $i$. We assume there are no self-loops, i.e.\ we have
$a_{ii}=0$. A path from node $i_1$ to $i_k$ is a sequence of
nodes $\{i_1,\ldots, i_k\}$ such that $(i_j, i_{j+1})\in \mathcal{E}$
for $j=1,\ldots, k-1$. A {directed tree} is a subgraph (subset of
nodes and edges) in which every node has exactly one parent node
except for one node, called the {root}, which has no parent
node. A {directed spanning tree} is a subgraph which is a
directed tree containing all the nodes of the original graph. If a
directed spanning tree exists, the root has a directed path to every
other node in the graph \cite{royle-godsil}.

The \emph{weighted in-degree} of a vertex $i$ is given by
\[
d_{\text{in}}(i) = \sum_{j=1}^N\, a_{ij}.
\]
For a weighted graph $\mathcal{G}$, the matrix
$L=[\ell_{ij}]$ with
\[
\ell_{ij}=
\begin{system}{cl}
	\sum_{k=1}^{N} a_{ik}, & i=j,\\
	-a_{ij}, & i\neq j,
\end{system}
\]
is called the {Laplacian matrix} associated with the graph
$\mathcal{G}$. The Laplacian matrix $L$ has all its eigenvalues in the
closed right half plane and at least one eigenvalue at zero associated
with right eigenvector $\textbf{1}$ \cite{royle-godsil}. Moreover, if
the graph contains a directed spanning tree, the Laplacian matrix $L$
has a single (simple) eigenvalue at the origin and all other eigenvalues are
located in the open right-half complex plane \cite{ren-book}. 

\begin{figure}[ht]
	\includegraphics[width=8cm]{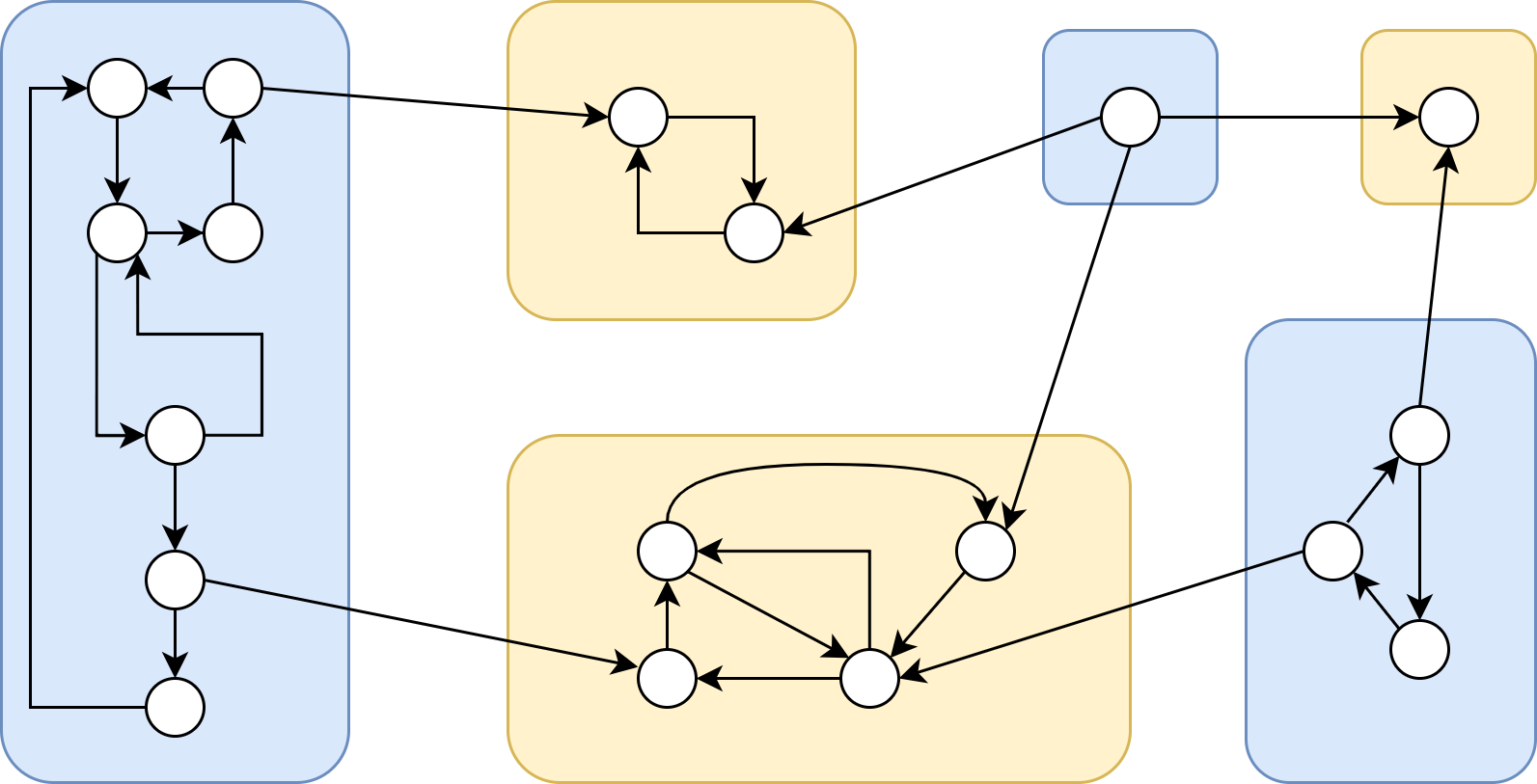}
	\centering
	\caption{A directed communication network and its bicomponents.}\label{f1}
\end{figure}

A directed communication network is said to be strongly connected if
it contains a directed path from every node to every other node in the
graph. For a given graph $\mathcal{G}$ every maximal (by inclusion)
strongly connected subgraph is called a bicomponent of the graph. A
bicomponent without any incoming edges is called a basic
bicomponent. Every graph has at least one basic bicomponent. Networks
have one unique basic bicomponent if and only if the network contains
a directed spanning tree. In general, every node in a network can be
reached from at least one basic bicomponent, see \cite[page
7]{stanoev-smilkov-2013}. In Fig. \ref{f1} a directed communication
network with its bicomponents is shown. The network in this figure
contains 6 bicomponents, 3 basic bicomponents (the blue ones) and 3
non-basic bicomponents (the yellow ones).

In the absence of a directed spanning tree, the Laplacian matrix of
the graph has an eigenvalue at the origin with a multiplicity $k$
larger than $1$. This implies that it is a $k$-reducible matrix and
the graph has $k$ basic bicomponents.  The book \cite[Definition
2.19]{wu-book} shows that, after a suitable permutation of the nodes,
a Laplacian matrix with $k$ basic bicomponents can be written in the
following form:
\begin{equation}\label{Lstruc}
	L=\begin{pmatrix}
		L_0 & L_{0,1}     & L_{0,2} & \cdots  & L_{0,k} \\
		0   & L_1        & 0      &  \cdots  & 0 \\
		\vdots & \ddots  & \ddots & \ddots  & \vdots \\
		\vdots &         & \ddots & L_{k-1} & 0 \\
		0      & \cdots & \cdots  & 0       & L_k
	\end{pmatrix}
\end{equation}
where $L_1,\ldots, L_k$ are the Laplacian matrices associated to the
$k$ basic bicomponents in our network. These matrices have a simple
eigenvalue in $0$ because they are associated with a strongly
connected component. On the other hand, $L_0$ contains all non-basic
bicomponents and is a grounded Laplacian with all eigenvalues in the
open right-half plane. 

\section{Scale-free non-collaborative protocol design for multi-agent systems}

Consider multi-agent systems (MAS) consisting of $N$ identical agents: 
\begin{equation}\label{eq1}
	\begin{system*}{ccl}
		x_i^+(t) &=& Ax_i(t)+Bu_i(t),\\
		y_i(t) &=& Cx_i(t),
	\end{system*}
\end{equation}
where $x_i(t)\in\mathbb{R}^{n}$, $y_i(t)\in\mathbb{R}$ and
$u_i(t)\in\mathbb{R}$ are the state, output, and input of agent $i$,
respectively, with $i=1,\ldots, N$. In the aforementioned
presentation, for continuous-time systems, $x_i^+(t) = \dot{x}_i(t)$
with $t \in \mathbb{R}$ while for discrete-time systems,
$x_i^+(t) = x_i(t + 1)$ with $t \in \mathbb{Z}$.

For continuous time MAS, the communication network provides agent $i$
with the following information,
\begin{equation}\label{zeta1}
	\zeta_i(t)=\sum_{j=1}^{N}a_{ij}(y_i(t)-y_j(t)).
\end{equation}
The communication topology of the
network can be described by a weighted and directed graph
$\mathcal{G}$ with nodes corresponding to the agents in the network
and the weight of edges given by coefficient $a_{ij}$. In terms of the
coefficients of the associated Laplacian matrix $L$, $\zeta_i(t)$ can
be rewritten as
\begin{equation}\label{zeta}
	\zeta_i(t)= \sum_{j=1}^{N}\ell_{ij}y_j(t).
\end{equation}

For discrete-time agents, each agent
$i\in\{1,\ldots,N\}$ has access to the quantity
\begin{equation}\label{zeta-dis}
	\zeta_i(t)=\dfrac{1}{1+q_i}\sum_{j=1}^N a_{ij}(y_i(t)-y_j(t)),
\end{equation}
where $q_i$ is an upper bound on $d_{\text{in}}(i)$ for
$i, j \in \{1,\ldots,N\}$ \cite{stoorvogel-saberi-liu}.  In that case, we can use the modified
information-exchange
\begin{equation}\label{zeta-y}
	\zeta_i(t)=\sum_{j=1,j\neq i}^N d_{ij}(y_i(t)-y_j(t)),
\end{equation}
instead of \eqref{zeta} where
\[
d_{ij}= \frac{a_{ij}}{1+q_i},
\]
for $i\neq j$ while 
\[
d_{ii}=1-\sum_{j=1,j\neq i}^Nd_{ij}.
\]
Note that the weight matrix $D=[d_{ij}]$ is then a, so-called, row
stochastic matrix.  Let 
\[
Q_{\text{in}}=\diag\left(q_1,q_2,\ldots,q_N\right).
\]
Then, the relationship between the
row stochastic matrix $D$ and the Laplacian matrix $L$ is
\begin{equation}\label{hodt-LD-dis}
	(I+Q_{\text{in}})^{-1}L=I-D.
\end{equation}

Non-collaborative protocols only use this relative measurement
$\zeta_i$ and achieve fully distributed protocols. In scale-free
protocols, we are looking for protocols which do not depend on the
network structure. This is motivated by the fact that in many
applications, an agent might be added/removed or a link might fail and
you then do not want to have to redesign the protocols being used.

\begin{definition}
	We denote by $\mathbb{G}^N$ the set of all directed graphs with $N$
	nodes which contain a directed spanning tree.
\end{definition}

We formulate the scale-free non-collaborative synchronization problem of a
MAS as follows.

\begin{problem}\label{prob4}
	The \textbf{scale-free non-collaborative state synchronization
		problem} for MAS \eqref{eq1} with communication given by
	\eqref{zeta} for continuous-time or \eqref{zeta-y} for discrete-time
	case is to find, if possible, a fixed linear protocol of the form:
	\begin{equation}\label{protoco1}
		\begin{system}{cl}
			x_{i,c}^+ &=A_{c}x_{i,c}+B_{c}\zeta_i,\\
			u_i &=F_{c}x_{i,c}+G_{c}\zeta_i,
		\end{system}
	\end{equation}
	where $x_{c,i}(t)\in\R^{n_c}$ is the state of protocol, such that
	state synchronization is achieved
	\begin{equation}\label{synch_state}
		\lim_{t\rightarrow \infty} x_i(t)-x_j(t) =0
	\end{equation}
	for all $i,j=1,\ldots,N$ for any number of agents $N$, for any fixed
	communication graph $\mathcal{G}\in\mathbb{G}^N$ and for all initial
	conditions of agents and protocols.
\end{problem}

We refer to a protocol \eqref{protoco1} which solves Problem
\ref{prob4} as a scale-free non-collaborative linear protocol.

\subsection{Continuous-time MAS}

In this section, we focus on continuous-time MAS. It is known that
the Problem \ref{prob4} is solvable for a large class of systems. We
recall the following theorem from \cite[Theorem
1]{liu-saberi-stoorvogel-tac-2023}.

\begin{theorem}\label{theorem1}
	The scale-free continuous-time state synchronization problem as
	formulated in Problem \ref{prob4} is solvable if the agent model
	\eqref{eq1} is either asymptotically stable or satisfies the
	following conditions:
	\begin{itemize}
		\item Stabilizable and detectable,
		\item Neutrally stable,
		\item Minimum phase,
		\item Uniform rank with the order of the infinite zero
		equal to one.
	\end{itemize}
\end{theorem}

This paper wants to investigate what happens if we apply a protocol of
the form \eqref{protoco1} designed to solve Problem \ref{prob4} to a
network which does \textbf{not} contain a directed spanning tree.

By using protocol \eqref{protoco1} we can define 
\begin{equation}\label{ABCtilde}
	\tilde{A}=\begin{pmatrix} A & BF_c \\ 0 & A_c \end{pmatrix},\quad
	\tilde{B}=\begin{pmatrix} BG_c \\ B_c \end{pmatrix},\quad
	\tilde{C}=\begin{pmatrix} C & 0 \end{pmatrix}.
\end{equation}
We have the following result:

\begin{theorem}\label{theorem2}
	Consider a continuous-time MAS with agent dynamics
	\eqref{eq1}. Assume a protocol \eqref{protoco1} solves the
	scale-free state synchronization problem, i.e.\ the protocol is
	designed for the case when  the network contains a
	directed spanning tree.
	
	If the network does not contain a directed spanning tree, then the
	graph contains $k$ basic bicomponents with $k>1$. Then for any
	$i\in\{ 1,\ldots,k\}$,
	\begin{itemize}
		\item Within basic bicomponent $\mathcal{B}_i$, the state of the
		agents and the state of the associated protocol achieve
		synchronization and converge to trajectories $x_{i,s}$ and
		$x_{i,s,c}$ respectively satisfying
		\[
		\begin{pmatrix} \dot{x}_{i,s} \\
			\dot{x}_{i,s,c} \end{pmatrix}=\tilde{A} \begin{pmatrix}
			x_{i,s} \\ x_{i,s,c} \end{pmatrix} 
		\]
		whose initial condition is a linear combination of the initial
		conditions of the agents within this basic bicomponent.
		\item An agent $j$ which is not part of any of the basic
		bicomponents synchronizes to a trajectory:
		\[
		\sum_{i=1}^k\, \beta_{j,i} \begin{pmatrix} x_{i,s} \\ x_{i,s,c}
		\end{pmatrix}
		\]
		where the coefficients $\beta_{j,i}$ are nonnegative, satisfy:
		\begin{equation}\label{betasum}
			1=\sum_{i=1}^k\, \beta_{j,i} 
		\end{equation}
		and only depend on the parameters of the network and do not depend
		on any of the initial conditions.
	\end{itemize}
\end{theorem}

\begin{proof}
	The paper \cite[Chapter 2]{saberi-stoorvogel-zhang-sannuti} has
	shown that we achieve scale-free state synchronization if and only
	if
	\begin{equation}\label{abc}
		\tilde{A}+\lambda \tilde{B}\tilde{C}
	\end{equation}
	is asymptotically stable for all $\lambda\in \C$ with
	$\re \lambda >0$. We use the decomposition \eqref{Lstruc} introduced
	before.  We label the states of the agents and protocols as:
	\[
	\bar{x}_i = \begin{pmatrix} x_{k_i+1} \\ x_{k_i+1,c}
		\\ x_{k_i+2} \\ x_{k_i+2,c} \\ \vdots \\ x_{k_{i+1}} \\ x_{k_{i+1},c}
	\end{pmatrix}
	\]
	where $k_0=0$ and $k_{i+1}-k_i$ is the number of columns of $L_i$
	for $i=0,\ldots,k$.  It is easy to verify that the dynamics of the
	$i$'th basic bicomponent ($i \in \{ 1,\ldots, k\}$) is given by:
	\[
	\dot{\bar{x}}_i = \left[ (I\otimes \tilde{A}) + L_i \otimes
	\tilde{B}\tilde{C} \right] \bar{x}_i
	\]
	and since \eqref{abc} is asymptotically stable for all
	$\lambda\in \C$ with $\re \lambda >0$, classical results guarantee
	synchronization within this basic bicomponent and
	\[
	\begin{pmatrix} x_{j}(t) \\ x_{j,c}(t)
	\end{pmatrix} - \begin{pmatrix} x_{i,s}(t) \\ x_{i,s,c}(t) \end{pmatrix}
	\rightarrow 0
	\]
	as $t\rightarrow \infty$ for $j=k_i+1,\ldots, k_{i+1}$ with:
	\[
	\begin{pmatrix} \dot{x}_{i,s} \\ \dot{x}_{i,s,c} \end{pmatrix}
	= \tilde{A} \begin{pmatrix} x_{i,s} \\ x_{i,s,c} \end{pmatrix}
	\]
	and
	\[
	\begin{pmatrix} x_{i,s}(0) \\ x_{i,s,c}(0) \end{pmatrix} =
	\sum_{i=k_i+1}^{k_{i+1}} \alpha_i \begin{pmatrix} x_{k_i+1}(0) \\ x_{k_i+1,c}(0)
	\end{pmatrix}
	\]
	where
	\[
	\bar{\alpha}_{i} = \begin{pmatrix} \alpha_{k_i+1} & \cdots &
		\alpha_{k_{i+1}} \end{pmatrix}
	\]
	is the unique left-eigenvector associated with eigenvalue $0$ of
	$L_i$ whose elements are nonnegative and sum up to $1$.  Define
	\begin{equation}\label{ererrr2}
		\bar{x}_{0,s} =-\sum_{i=1}^k (L_0^{-1}L_{0,i}\textbf{1}_i \otimes
		I) \begin{pmatrix} x_{i,s} \\ x_{i,s,c} \end{pmatrix}
	\end{equation}
	where $\textbf{1}_i$ is a column vector with all elements equal to
	$1$ whose size is equal to the number of columns of $L_{0,i}$ while
	$\textbf{1}_0$ is a column vector with all elements equal to $1$
	whose size is equal to the number of columns of $L_0$. We claim that
	$\bar{x}_0 (t)-\bar{x}_{0,s}(t) \rightarrow 0$ as
	$t\rightarrow \infty$ which implies that
	\begin{equation}\label{beta1}  
		\begin{pmatrix} x_{j}(t) \\ x_{j,c}(t)
		\end{pmatrix} - \sum_{i=1}^k \beta_{j,i} \begin{pmatrix}
			x_{i,s}(t) \\ x_{i,s,c}(t) \end{pmatrix} \rightarrow 0
	\end{equation}
	as $t\rightarrow \infty$ for $j=1,\ldots, k_1$ where
	\begin{equation}\label{beta2}
		\begin{pmatrix} \beta_{1,i} \\ \beta_{2,i} \\ \vdots \\
			\beta_{k_0,i} \end{pmatrix} = -L_0^{-1}L_{0,i}\textbf{1}_i.
	\end{equation}
	Note that the elements of $L_{0,i}$ are all nonpositive and the
	elements of $L_0^{-1}$ are all nonnegative (as the inverse of a
	grounded Laplacian matrix). This immediately shows that the
	coefficients $\beta_{j,i}$ are all nonnegative. Moreover,
	\begin{equation}\label{star3}
		L\begin{pmatrix} \textbf{1}_0 \\ \vdots \\
			\textbf{1}_k \end{pmatrix} = 0
	\end{equation}
	implies
	\begin{equation}\label{ererrr}
		L_0\textbf{1}_0 + \sum_{i=1}^k L_{0,i} \textbf{1}_i = 0.
	\end{equation}
	Note that \eqref{ererrr} implies that the $\beta_{i,j}$ defined in
	\eqref{beta2} satisfies \eqref{betasum}. Remains to verify that
	\eqref{beta1} is satisfied.  In order to establish this we look at
	the dynamics of the agents not contained in one of the basic
	bicomponents. We get:
	\[
	\dot{\bar{x}}_0 = \left[ (I\otimes \tilde{A}) + L_0 \otimes
	\tilde{B}\tilde{C} \right] \bar{x}_0
	+ \sum_{i=1}^k  (L_{0,i} \otimes
	\tilde{B}\tilde{C} ) \bar{x}_i.
	\]
	After some algebraic manipulations we find that
	\[
	e_0 = \bar{x}_0-\bar{x}_{0,s}
	\]
	satisfies:
	\begin{multline*}
		\dot{e}_0 = \left[ (I\otimes \tilde{A}) + L_0 \otimes
		\tilde{B}\tilde{C} \right] e_0 \\+
		\sum_{i=1}^k (L_{0,i}\otimes \tilde{B}\tilde{C})
		\left[\bar{x}_i-(\textbf{1}_i\otimes I)\begin{pmatrix} x_{i,s} \\
			x_{i,s,c} \end{pmatrix}\right] 
	\end{multline*}
	where we use that \eqref{ererrr2} implies:
	\[
	(L_0\otimes \tilde{B}\tilde{C}) \bar{x}_{0,s} +
	\sum_{i=1}^k (L_{0,i} \textbf{1}_i\otimes \tilde{B}\tilde{C})\begin{pmatrix} x_{i,s} \\
		x_{i,s,c} \end{pmatrix} = 0
	\]
	
	Since $I\otimes \tilde{A} + L_0 \otimes \tilde{B}\tilde{C}$ is
	asymptotically stable and
	\[
	\bar{x}_i(t)-(\textbf{1}_i\otimes I)\begin{pmatrix} x_{i,s}(t) \\
		x_{i,s,c}(t) \end{pmatrix}
	\rightarrow 0
	\]
	as $t\rightarrow \infty$ we find that $e_0(t)\rightarrow 0$ as
	$t\rightarrow \infty$ which yields \eqref{beta1}.
\end{proof}

\begin{remark}
	In the above proof, we established the synchronization properties
	for a general network structure. Let's investigate how the above is
	consistent with the result that we achieve synchronization in case
	the network has a directed spanning tree.
	
	The network having a directed spanning tree is equivalent to the
	network having a single basic bicomponent. In the notation of the
	above proof we then have $k=1$. We note that the state of the agents
	and the state of their associated protocol within the single basic
	bicomponent converge to some trajectory:
	\[
	\begin{pmatrix}
		x_{1,s}(t) \\ x_{1,s,c}(t) \end{pmatrix}
	\]
	Next we note that \eqref{ererrr} with $k=1$ implies
	$\textbf{\textup{1}}_0 =-L_0^{-1}
	L_{0,1}\textbf{\textup{1}}_1$. This yields that \eqref{beta2}
	reduces to:
	\[
	\begin{pmatrix} \beta_{1,1} \\ \beta_{2,1} \\ \vdots \\
		\beta_{k_0,1} \end{pmatrix} = -L_0^{-1}L_{0,1}\textbf{\textup{1}}_1
	=\textbf{\textup{1}}_0
	\]
	which implies that \eqref{beta1} reduces to:
	\[
	\begin{pmatrix} x_{j}(t) \\ x_{j,c}(t)
	\end{pmatrix} -  \begin{pmatrix}
		x_{1,s}(t) \\ x_{1,s,c}(t) \end{pmatrix} \rightarrow 0
	\]
	as $t\rightarrow \infty$ for $j=1,\ldots, k_1$. This implies that we
	indeed achieve synchronization.
\end{remark}

\subsection{Discrete-time MAS}

We focus on the discrete-time MAS.  The following solvability results
are known by recalling \cite[Theorem
2]{liu-saberi-stoorvogel-tac-2023}.

\begin{theorem}\label{theorem1d}
	The scale-free discrete-time non-collaborative state synchronization
	problem as formulated in Problem \ref{prob4} is solvable with
	information exchange \eqref{zeta-y} if the agent model \eqref{eq1}
	is either asymptotically stable or satisfies the following
	conditions:
	\begin{itemize}
		\item Stabilizable and detectable,
		\item Neutrally stable.
	\end{itemize}
\end{theorem}

Note that in the discrete-time case we do not need restrictions on the
zeros which is actually due to our use of local bounds on the
neighborhoods that yielded the modified exchange \eqref{zeta-y}. The
following result shows that in discrete time we effectively get the
same result as in continuous time.

\begin{theorem}\label{theorem2d}
	Consider a discrete-time MAS with agent dynamics \eqref{eq1}. Assume
	a protocol \eqref{protoco1} solves the scale-free non-collaborative
	state synchronization problem, i.e.\ the protocol is
	designed for the case when  the network contains a
	directed spanning tree.
	
	If the network does not contain a directed spanning tree, then the
	Laplacian matrix of the graph has an eigenvalue at the origin with a
	multiplicity $k$ larger than $1$. This implies that the graph has $k$
	basic bicomponents. Then for any $i\in(1,\ldots,k)$,
	\begin{itemize}
		\item Within basic bicomponent $\mathcal{B}_i$, the state of the
		agents and the state of the associated protocol achieve
		synchronization and converge to trajectories $x_{i,s}$ and
		$x_{i,s,c}$ respectively satisfying
		\[
		\begin{pmatrix} {x}_{i,s}(t+1) \\
			x_{i,s,c}(t+1)  \end{pmatrix}=\tilde{A} \begin{pmatrix}
			x_{i,s}(t)  \\ x_{i,s,c}(t)  \end{pmatrix} 
		\]
		whose initial condition is a linear combination of the initial
		conditions of the agents within this basic bicomponent.
		\item An agent $j$ which is not part of any of the basic
		bicomponents synchronizes to a trajectory:
		\[
		\sum_{i=1}^k\, \beta_{j,i} \begin{pmatrix} x_{i,s}(t) \\ x_{i,s,c}(t)
		\end{pmatrix}
		\]
		where the coefficients $\beta_{j,i}$ are nonnegative, satisfy
		\eqref{betasum} and only depend on the parameters of the network
		and do not depend on any of the initial conditions.
	\end{itemize}
\end{theorem}

\begin{proof}
	The proof of Theorem \ref{theorem2} can be easily modified to yield
	the above result. For instance, instead of \eqref{abc} we need that
	\begin{equation}\label{abc2}
		\tilde{A}+(1-\lambda) \tilde{B}\tilde{C}
	\end{equation}
	for all $\lambda$ in the open unit disc and the structure of the
	matrix $L$ immediately relates to a similar structure of the row
	stochastic matrix $D$.
\end{proof}

\section{Scale-free collaborative protocol design for multi-agent systems}

Consider the same class of multi-agent systems (MAS) consisting of $N$
identical agents of the form \eqref{eq1} with information exchange
given by \eqref{zeta}.

Collaborative protocols, which were introduced by
\cite{li-duan-chen-huang}, allow extra information exchange between
neighbors. Typically, this additional information exchange consists of
relative information about the difference between the state of the
protocol of a specific agent and the state of the protocol of a
neighboring agent using the same network.  In other words, we also
have:
\begin{equation}\label{zeta2}
	\hat{\zeta}_i = \sum_{j=1}^{N}\ell_{ij} H_c x_{i,c}
\end{equation}
available in our protocol design where $x_{i,c}$ denotes the state of
the protocol for the $i$'th agent and the matrix $H_c$ can be part of
the protocol design.

For discrete-time MAS, we have
\begin{equation}\label{zeta2-d}
	\hat{\zeta}_i (t)= \sum_{j=1,j\neq i}^{N}d_{ij} H_c (x_{i,c}(t)-x_{j,c}(t))
\end{equation}

We formulate the scale-free collaborative synchronization problem of a
MAS as follows.

\begin{problem}\label{prob5}
	The \textbf{scale-free collaborative state synchronization problem}
	for MAS \eqref{eq1} with communication given by \eqref{zeta} and
	\eqref{zeta2} for continuous-time or \eqref{zeta-y} and
	\eqref{zeta2-d} for discrete-time is to find, if possible, a fixed
	linear protocol of the form:
	\begin{equation}\label{protoco2}
		\begin{system}{cl}
			x_{i,c}^+ &=A_c x_{i,c}+B_c \zeta_i + E_c \hat{\zeta}_i,\\
			u_i &=F_c x_{i,c}+ G^1_c \zeta_i + G^2_c \hat{\zeta}_i,
		\end{system}
	\end{equation}
	and a matrix $H_c$ where $x_{c,i}(t)\in\R^{n_c}$ is the state of
	protocol, such that state synchronization is achieved, i.e.\
	\eqref{synch_state} is satisfied for all $i,j=1,\ldots,N$ for any
	number of agents $N$, for any fixed communication graph
	$\mathcal{G}\in\mathbb{G}^N$ and for all initial conditions of
	agents and protocols.
\end{problem}

We call a protocol \eqref{protoco2} which solves Problem \ref{prob5},
a scale-free collaborative linear protocol.

\subsection{Continuous-time MAS}

Firstly, we obtain the necessary and sufficient conditions for
solvability of scale-free collaborative state synchronization for
continuous-time MAS. It is known that this problem is solvable under
some conditions. The main advantage of the collaborative protocols in
this context are that we no longer need restrictions on the zeros and
the relative degree of the system. Moreover, the requirement of
neutrally stability has been weakened to at most weakly unstable
condition.

\begin{theorem}\label{theorem3}
	The scale-free collaborative continuous-time state synchronization
	problem as formulated in Problem \ref{prob5} is solvable if and only
	if the agent model \eqref{eq1} is either asymptotically stable or
	satisfies the following conditions:
	\begin{itemize}
		\item Stabilizable and detectable,
		\item All poles are in the closed left-half plane.
	\end{itemize}
\end{theorem}

\begin{proof}
	Sufficiency has been established in the book \cite[Chapter
	3]{liu-nojavanzedah-saberi-2022-book} by explicitly constructing
	appropriate protocols.
	
	We only provide the proof of necessity. Stabilizabiliy and
	detectability are obviously necessary. By using protocol
	\eqref{protoco2} we can define
	\begin{equation}\label{ABCtilde2}
		\tilde{A}=\begin{pmatrix} A & BF_c \\ 0 & A_c \end{pmatrix},
		\tilde{B}=\begin{pmatrix} BG_c^1 & BG_c^2 \\ B_c & E_c \end{pmatrix},
		\tilde{C}=\begin{pmatrix} C & 0 \\ 0 & H_c \end{pmatrix}
	\end{equation}
	A continuous-time scale-free design requires:
	\[
	\tilde{A} + \lambda \tilde{B}\tilde{C}
	\]
	to be asymptotically stable for all $\lambda$ with
	$\re \lambda \geq 0$. Letting $\lambda\to 0$ we find as a necessary
	condition that the eigenvalues of $\tilde{A}$ must be in the closed
	left-half plane. This yields that all poles of the agents
	must be in the closed left-half plane.
\end{proof}

We again want to investigate what happens if we apply a protocol of
the form \eqref{protoco2} designed to solve Problem \ref{prob5} to a
network which does \textbf{not} contain a directed spanning tree. We
have the following result:

\begin{theorem}\label{theorem4}
	Consider a continuous-time MAS with agent dynamics \eqref{eq1} and
	communication via \eqref{zeta} and \eqref{zeta2}. Assume a protocol
	\eqref{protoco2} solves the scale-free state synchronization problem
	i.e.\ the protocol is
	designed for the case when  the network contains a
	directed spanning tree.
	
	If the network does not contain a directed spanning tree, then the
	Laplacian matrix of the graph has an eigenvalue at the origin with a
	multiplicity $k$ larger than $1$. This implies that the graph has $k$
	basic bicomponents. Then for any $i\in \{1,\ldots,k \}$,
	\begin{itemize}
		\item Within basic bicomponent $\mathcal{B}_i$, the state of the
		agents and the state of the associated protocol achieve
		synchronization and converge to trajectories $x_{i,s}$ and
		$x_{i,s,c}$ respectively satisfying
		\[
		\begin{pmatrix} \dot{x}_{i,s} \\
			\dot{x}_{i,s,c} \end{pmatrix}=\tilde{A} \begin{pmatrix}
			x_{i,s} \\ x_{i,s,c} \end{pmatrix} 
		\]
		whose initial condition is a linear combination of the initial
		conditions of the agents within this basic bicomponent.
		\item An agent $j$ which is not part of any of the basic
		bicomponents synchronizes to a trajectory:
		\[
		\sum_{i=1}^k\, \beta_{j,i} \begin{pmatrix} x_{i,s} \\ x_{i,s,c}
		\end{pmatrix}
		\]
		where the coefficients $\beta_{j,i}$ are nonnegative, satisfy
		\eqref{betasum} and only depend on the parameters of the network
		and do not depend on any of the initial conditions.
	\end{itemize}
\end{theorem}

\begin{remark}
	The main issue of the above theorem is that it shows that the extra
	communication does not have any influence on the synchronization
	properties of the network.
	
	We obtain the same synchronization properties using collaborative
	protocols as we obtained earlier for non-collaborative protocols.
\end{remark}

\begin{proof}
	The proof of Theorem \ref{theorem2} can be also used for this case
	with the only modification that \eqref{ABCtilde} is replaced by
	\eqref{ABCtilde2}.
\end{proof}

\subsection{Discrete-time MAS}

For discrete-time agents, we also have necessary and sufficient
conditions for solvability of scale-free collaborative state
synchronization as presented in the following result.

\begin{theorem}\label{theorem3-c}
	The scale-free collaborative discrete-time state synchronization
	problem as formulated in Problem \ref{prob5} is solvable by using
	protocol \eqref{protoco2} if and only if the agent model \eqref{eq1}
	is either asymptotically stable or satisfies the following
	conditions:
	\begin{itemize}
		\item Stabilizable and detectable,
		\item All poles are in the closed unit circle.
	\end{itemize}
\end{theorem}

\begin{proof}
	Sufficiency has been established in the book \cite[Chapter
	4]{liu-nojavanzedah-saberi-2022-book} by explicitly constructing
	appropriate protocols.
	
	We only provide the proof of necessity. By using protocol
	\eqref{protoco2} we still have \eqref{ABCtilde2}.  For the
	discrete-time MAS, a scale-free design requires:
	\[
	\tilde{A} + (1-\lambda) \tilde{B}\tilde{C}
	\]
	to be asymptotically stable for all $\lambda$ with
	$|\lambda| \leq 1$. Letting $\lambda \to 1$ we find as a necessary
	condition that the eigenvalues of $\tilde{A}$ must be in the closed
	unit circle. This immediately yields that all poles of the agents
	must be in the closed unit circle.
\end{proof}

Next, we again want to investigate what happens if we apply a protocol
of the form \eqref{protoco2} designed to solve Problem \ref{prob5} to
a network which does \textbf{not} contain a directed spanning tree:

\begin{theorem}\label{theorem4-d}
	Consider a discrete-time MAS with agent dynamics \eqref{eq1} and
	communication via \eqref{zeta-y} and \eqref{zeta2-d}. Assume a
	protocol \eqref{protoco2} solves the scale-free state
	synchronization problem, i.e.\ the protocol is
	designed for the case when  the network contains a
	directed spanning tree. We then call the protocol \eqref{protoco2} a
	scale-free collaborative linear protocol.
	
	If the network does not contain a directed spanning tree, then the
	Laplacian matrix of the graph has an eigenvalue at the origin with a
	multiplicity $k$ larger than $1$. This implies that the graph has $k$
	basic bicomponents. Then for any $i\in(1,\ldots,k)$,
	\begin{itemize}
		\item Within basic bicomponent $\mathcal{B}_i$, the state of the
		agents and the state of the associated protocol achieve
		synchronization and converge to trajectories $x_{i,s}$ and
		$x_{i,s,c}$ respectively satisfying
		\[
		\begin{pmatrix} {x}_{i,s}(t+1) \\
			x_{i,s,c}(t+1)  \end{pmatrix}=\tilde{A} \begin{pmatrix}
			x_{i,s}(t)  \\ x_{i,s,c}(t)  \end{pmatrix} 
		\]
		whose initial condition is a linear combination of the initial
		conditions of the agents within this basic bicomponent.
		\item An agent $j$ which is not part of any of the basic
		bicomponents synchronizes to a trajectory:
		\[
		\sum_{i=1}^k\, \beta_{j,i} \begin{pmatrix} x_{i,s}(t) \\ x_{i,s,c}(t)
		\end{pmatrix}
		\]
		where the coefficients $\beta_{j,i}$ are nonnegative, satisfy
		\eqref{betasum} and only depend on the parameters of the network
		and do not depend on any of the initial conditions.
	\end{itemize}
\end{theorem}

\begin{proof}
	Similar to the proof of Theorems \ref{theorem4} and \ref{theorem2},
	the proof can be also used for this case with the only modification
	that \eqref{ABCtilde} is replaced by \eqref{ABCtilde2}.
\end{proof}

\begin{remark}
	Note that in the conversion from a Laplacian matrix $L$ to a row
	stochastic matrix $D$ we used the local upper bounds $q_i$, see
	equation \eqref{zeta-dis}. It can be easily shown that the choice of
	$q_i$ does not affect the parameters $\beta_{j,i}$ in the above
	theorem. However, this choice can affect the initial conditions for
	$x_{i,s}$ and $x_{i,s,c}$ for the synchronized trajectory within
	basic bicomponent $\mathcal{B}_i$.
\end{remark}

\section{Numerical examples}

In this section, we show the efficiency of our protocol design by two
examples. We consider non-collaborative protocol
\eqref{protoco1}. We choose the existing examples in both continuous- and discrete-time presented in
\cite{liu-saberi-stoorvogel-tac-2023}.

\subsection{Continuous-time case}
Consider a continuous-time homogeneous MAS with agent model \eqref{eq1} given by:
\begin{equation*}
	A=\begin{pmatrix}
		0&1&1\\-1&0&1\\0&0&0
	\end{pmatrix},\  B=I,\ C=\begin{pmatrix}
		1&0&0\\
		0&1&0
	\end{pmatrix}.
\end{equation*}

Using the protocol design from
\cite{liu-saberi-stoorvogel-tac-2023}, we obtain the following non-collaborative protocols for $i=1,\dots,N$, 
\begin{equation}\label{pro-lin-partial-ex}
	\begin{system}{cl}
		\dot{z}_i&=\begin{pmatrix}
			-1&0&1\\-1&-2&1\\-1&-1&0
		\end{pmatrix}z_i+\begin{pmatrix}
			1&1\\0&2\\1&1
		\end{pmatrix}{\zeta}_i,\\
		\dot{p}_i&=-2p_i-\rho \begin{pmatrix}
			-0.6&-2.2&2.3&3.4
		\end{pmatrix}\begin{pmatrix}
		z_i\\\begin{pmatrix}
			0&1
		\end{pmatrix}{\zeta}_i
		\end{pmatrix},\\
		u_i&=-\begin{pmatrix}
			0\\0\\1
		\end{pmatrix}p_i-\rho \begin{pmatrix}
		0&0&0&0\\
		-0.6&-2.2&2.3&3.4\\
		0&0&0&0
		\end{pmatrix}\begin{pmatrix}
		z_i\\\begin{pmatrix}
			0&1
		\end{pmatrix}{\zeta}_i\end{pmatrix}.
	\end{system}
\end{equation}

\begin{figure}[ht]
	\includegraphics[width=8cm]{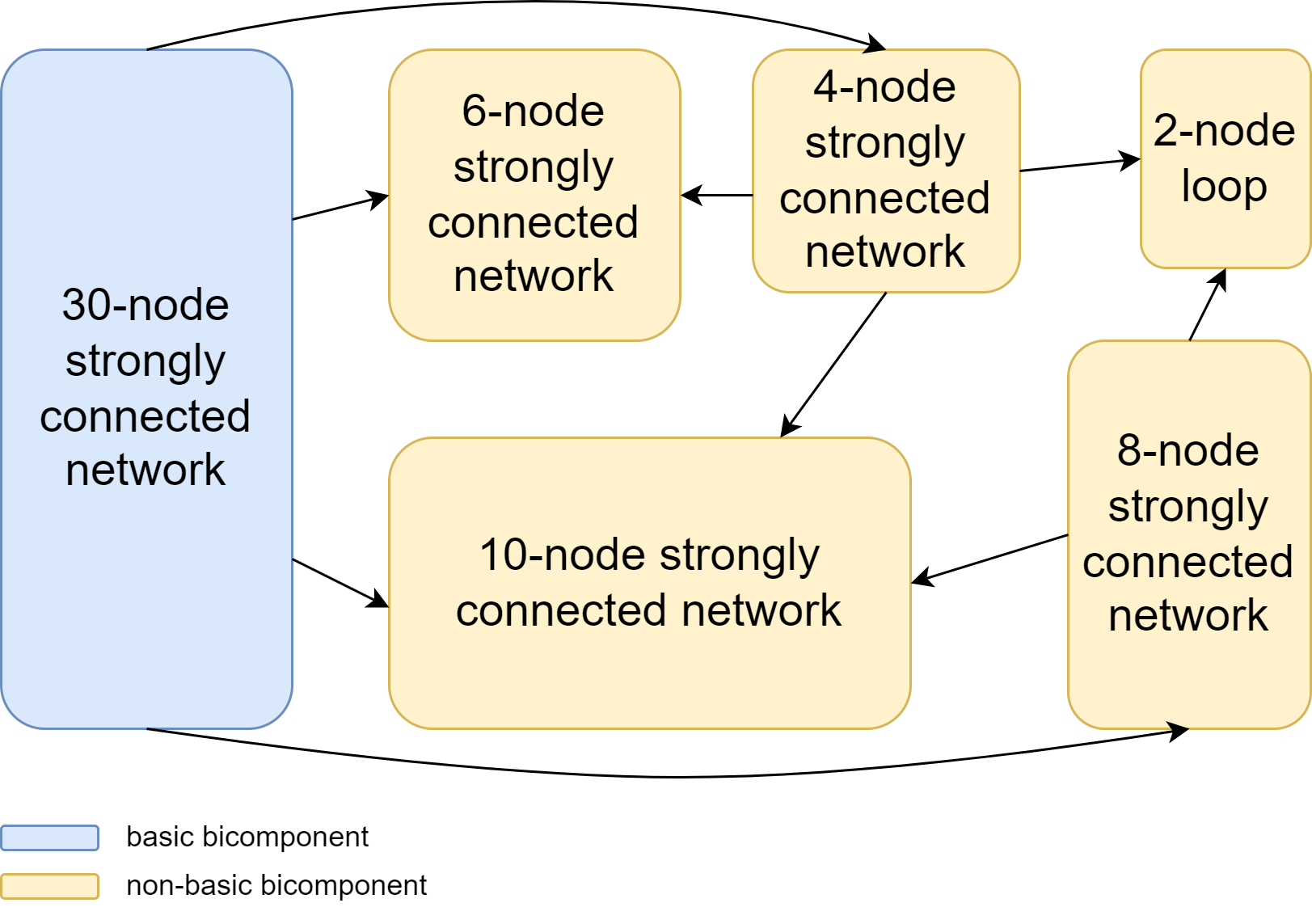}
	\centering
	\caption{The 60-nodes communication network with spanning tree.}\label{f5}
\end{figure}

We achieve output synchronization result for the 60-node heterogeneous
network shown in figure \ref{f5}, since it contains a directed
spanning tree and the protocol presented above is scale-free.

\begin{figure}[ht]
	\includegraphics[width=8cm]{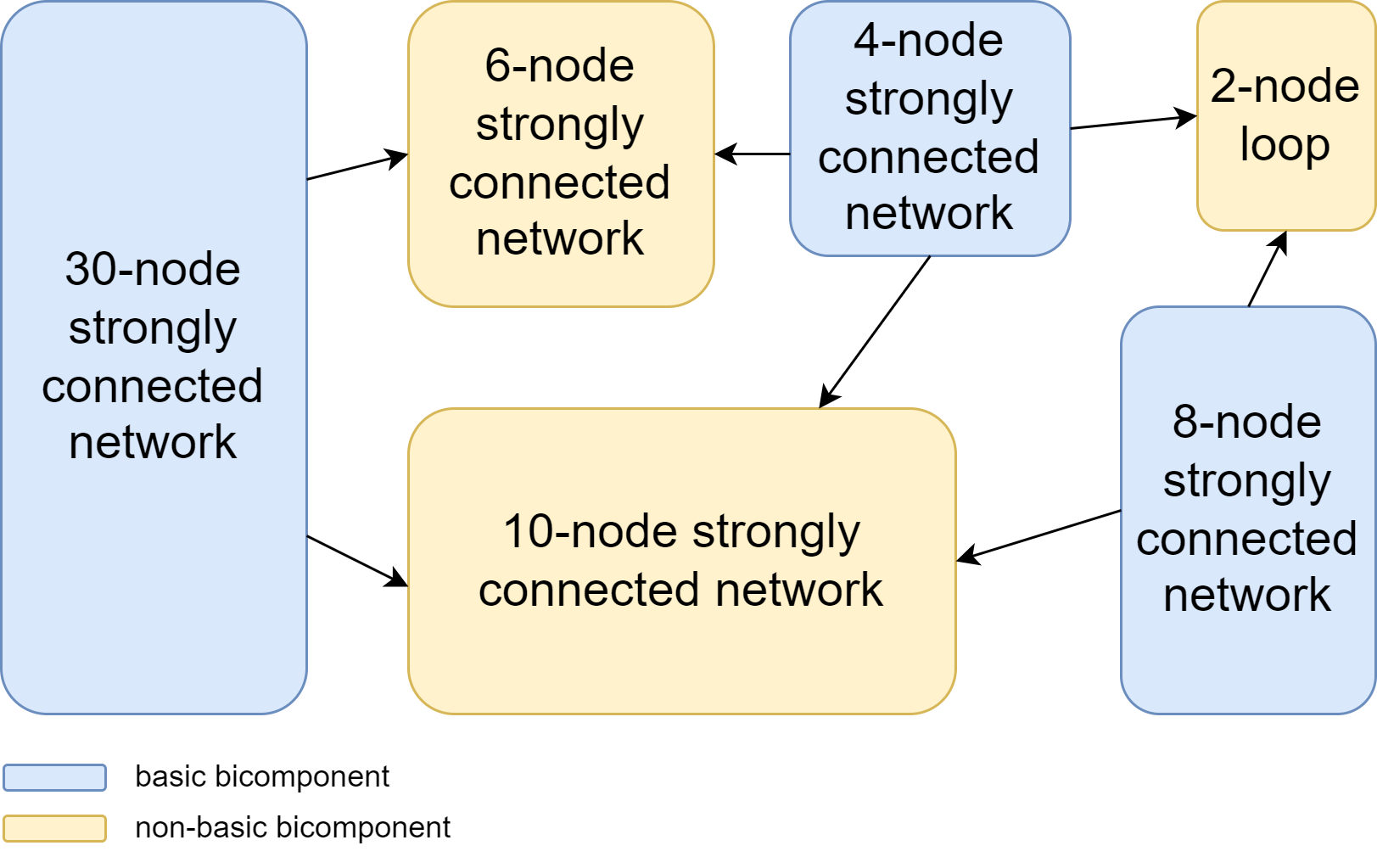}
	\centering
	\caption{The communication network without spanning tree. The
		links are broken due to faults. }\label{f4}  
\end{figure}

When some links have faults in the network, the communication network
might lose its directed spanning tree. For example, if two specific
links are broken in the original 60-node network given by
Fig. \ref{f5}, then we obtain the network as given in Figure \ref{f4}

It is obvious that there is no spanning tree in Figure \ref{f4}. We
obtain three basic bicomponents (indicated in blue): The 30-node, 8-node, and 4-node basic bicomponents. Meanwhile,
there are three non-basic bicomponents: a 10-node non-basic
bicomponent, a 6-node non-basic bicomponents and a 2-node non-basic bicomponents, which are indicated in
yellow.

%

\begin{figure}[p]
	\includegraphics[width=8cm]{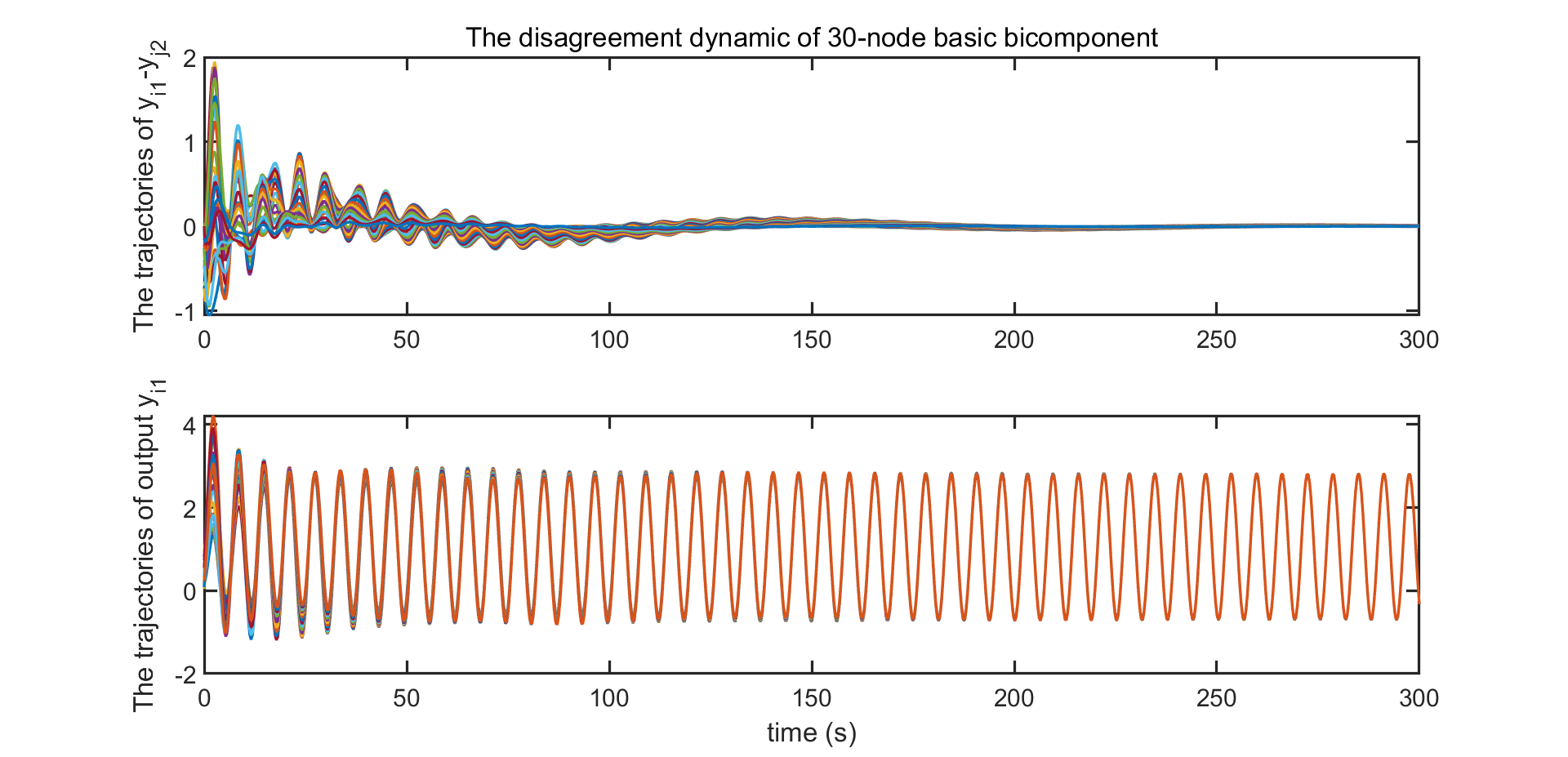}
	\includegraphics[width=8cm]{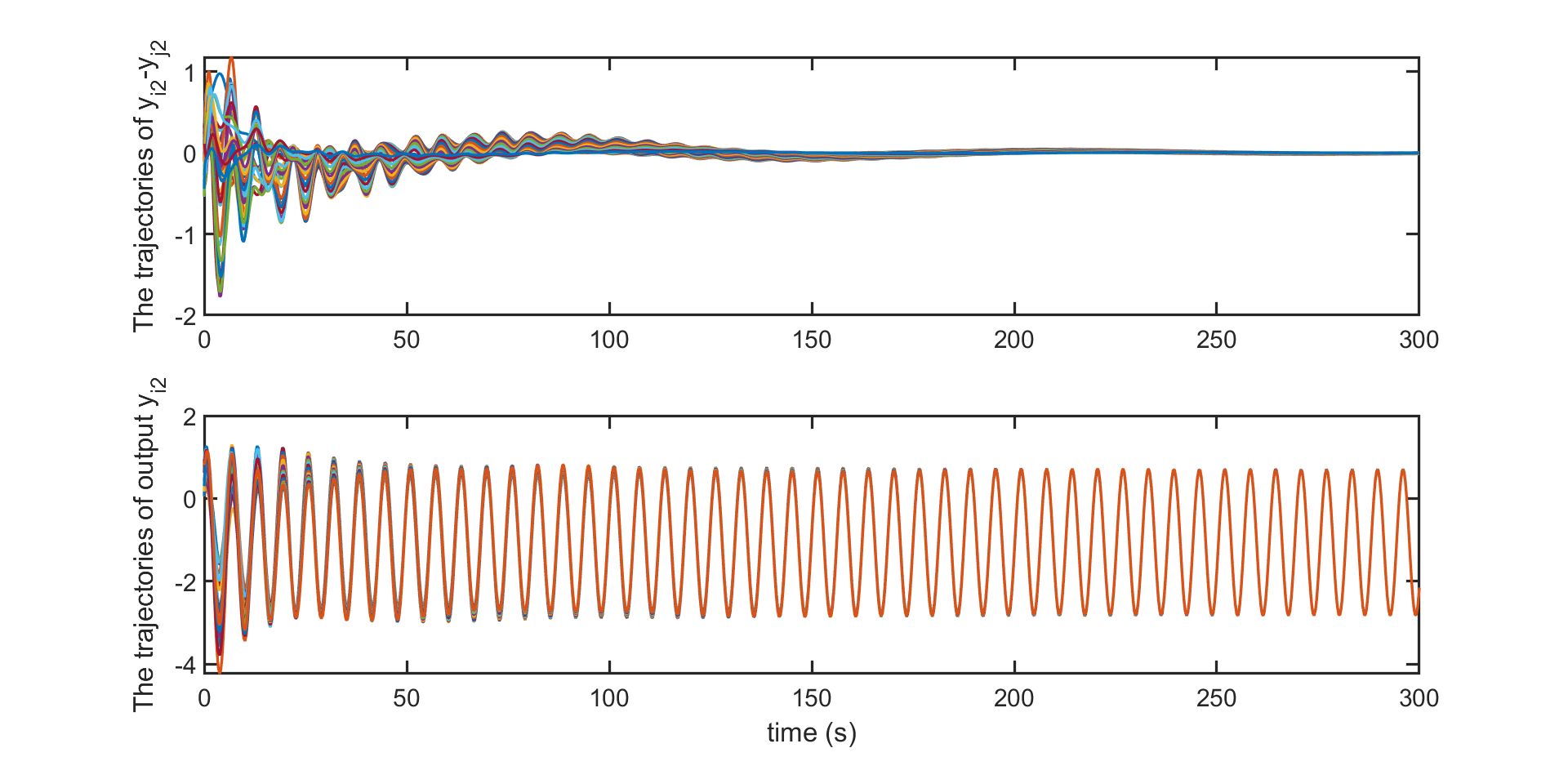}
	\centering
	\caption{30-node basic bicomponent for continuous-time MAS: disagreement dynamic among the
		agents and synchronized output trajectories.}\label{ct30cl}
\end{figure}

\begin{figure}[p]
	\includegraphics[width=8cm]{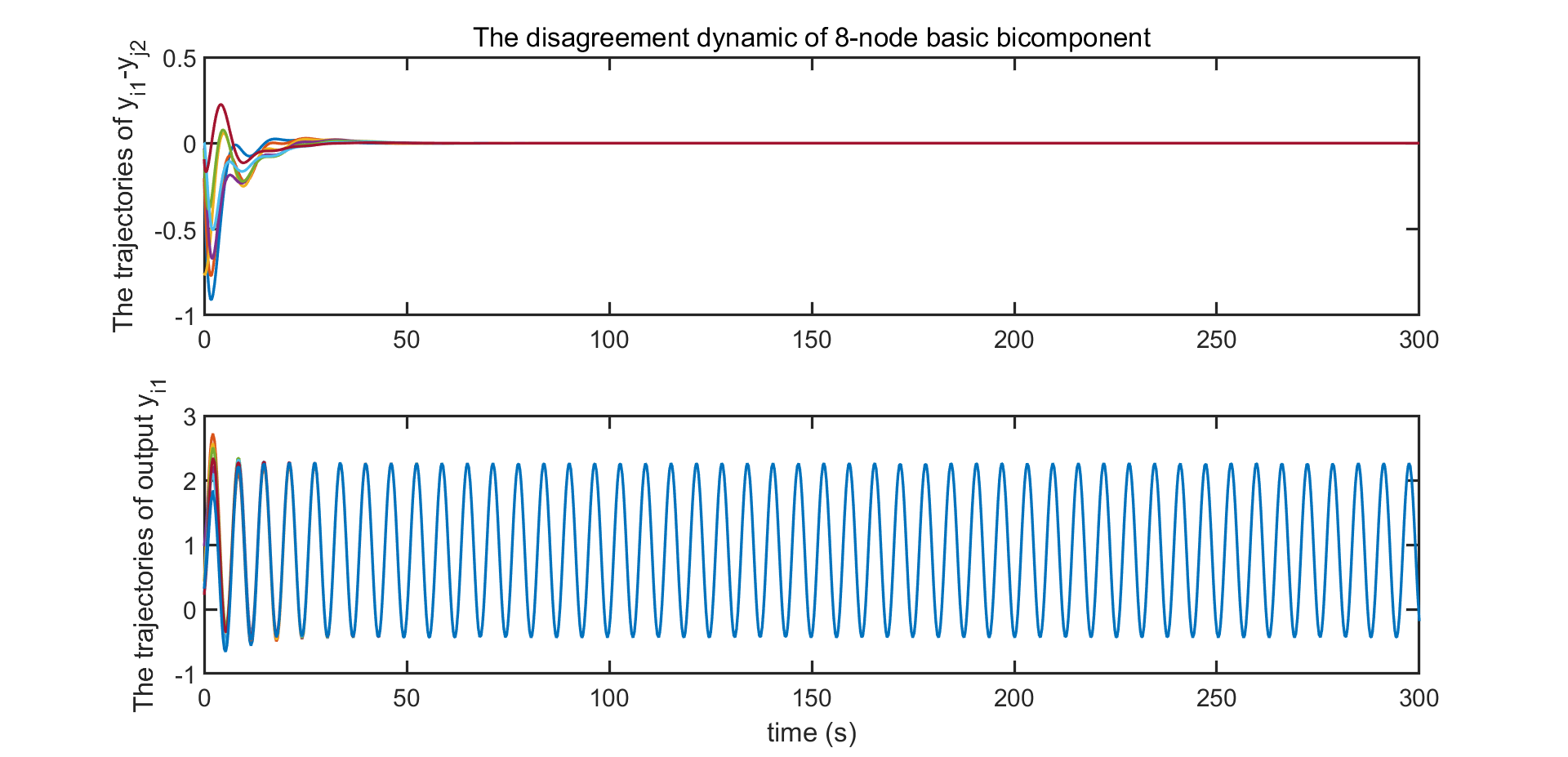} 
	\includegraphics[width=8cm]{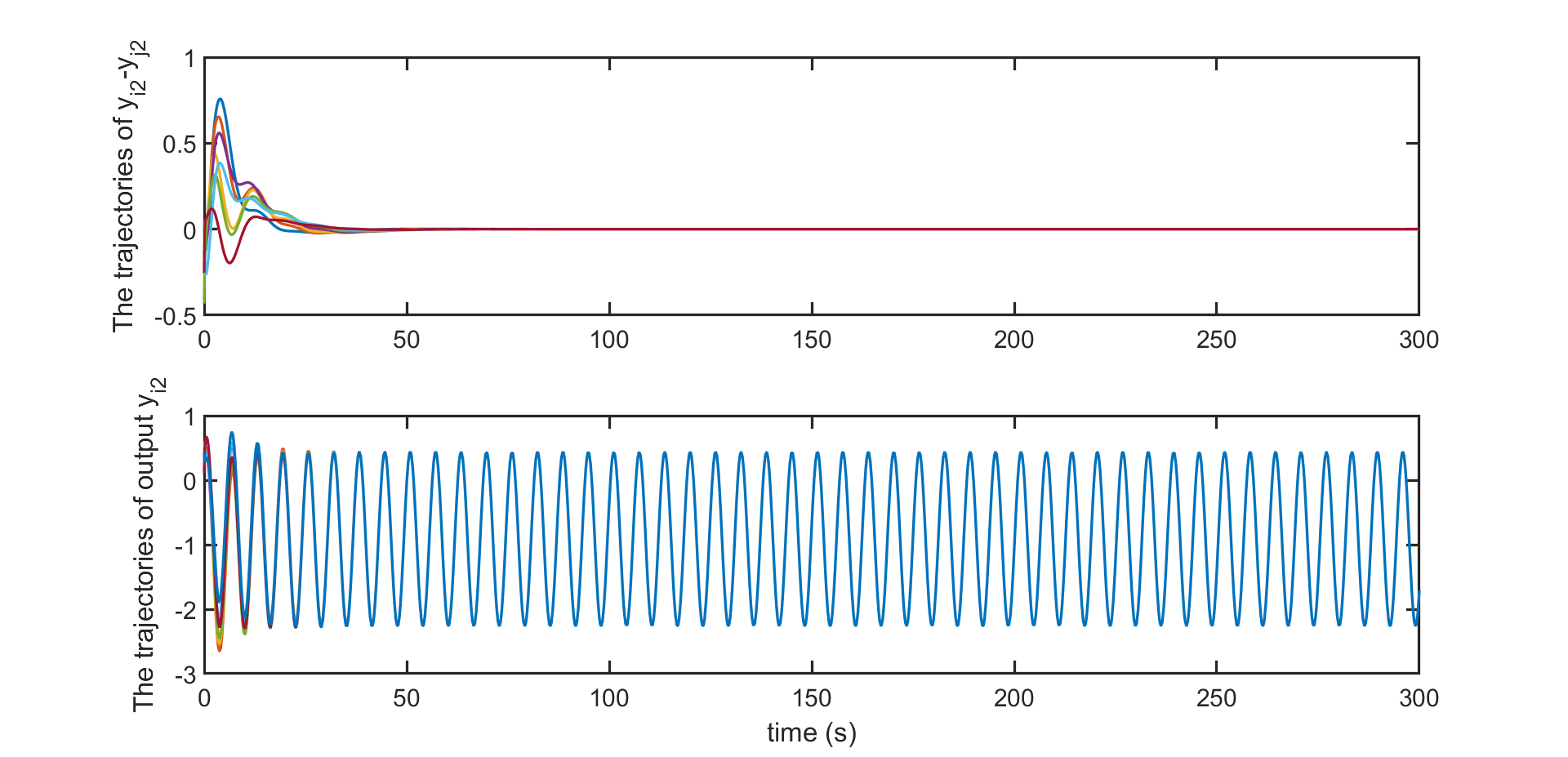}\centering
	\caption{8-node basic bicomponent for continuous-time MAS: disagreement dynamic among the
		agents and synchronized output trajectories.}\label{ct8cl}
\end{figure}

\begin{figure}[p]
	\includegraphics[width=9cm]{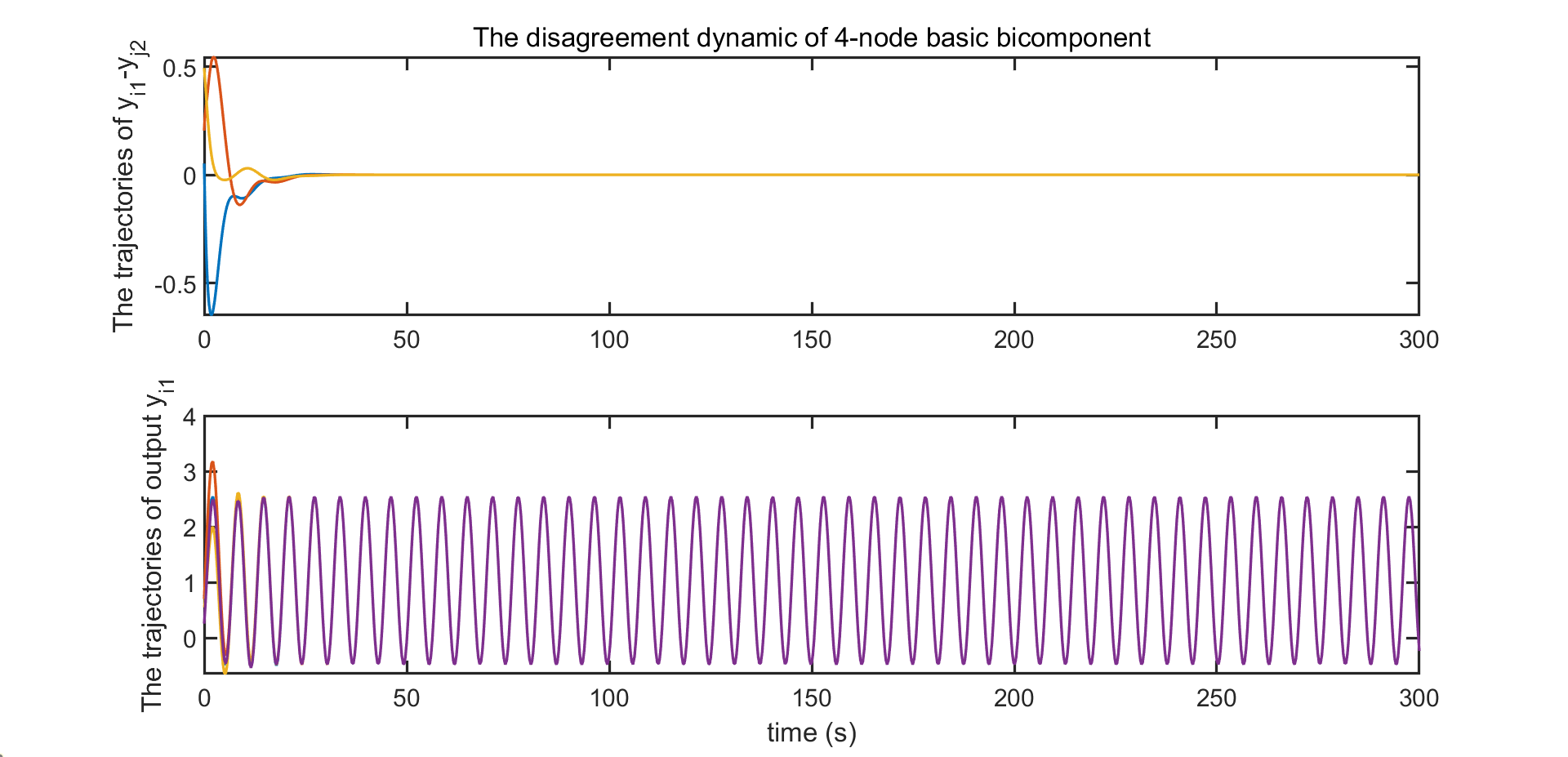}
	\includegraphics[width=9cm]{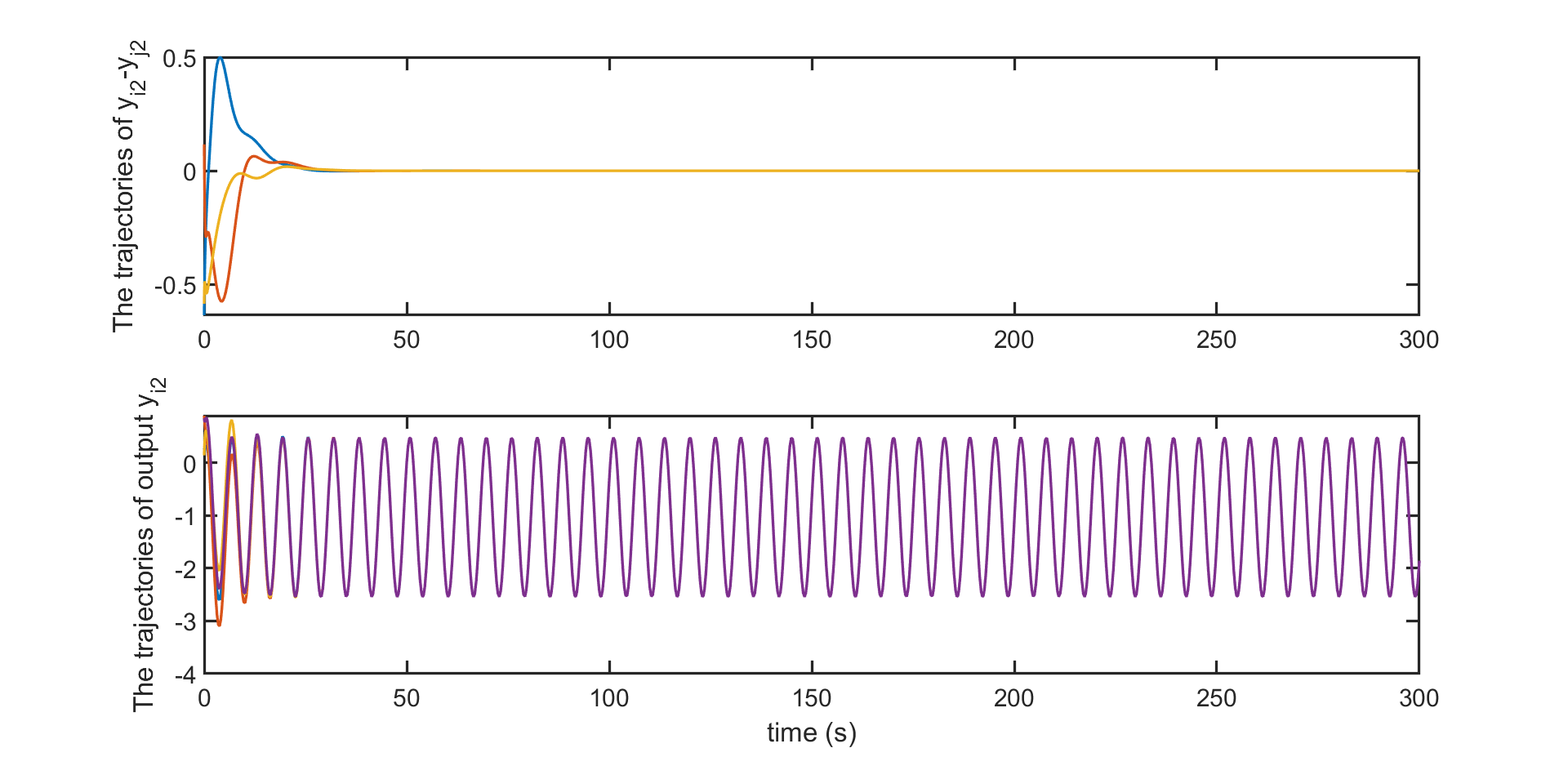}
	\centering
	\caption{4-node basic bicomponent for continuous-time MAS: disagreement dynamic among the
		agents and synchronized output trajectories.}\label{ct4cl}
\end{figure}

We have seen that for the 60-node network given by \eqref{f5} this
protocol indeed achieves output synchronization. If we apply the
protocol \eqref{pro-lin-partial-ex} to the network described by Figure
\ref{f4} which does not contain
a directed spanning tree, we again consider the six bicomponents
constituting the network. We see that, consistent with the theory, we
get output synchronization within figures \ref{ct30cl}, \ref{ct8cl}
and \ref{ct4cl} respectively which are the three basic
bicomponents. Obviously, the disagreement dynamic among the agents
(the errors between the output of agents) goes to zero within each
basic bicomponent. According to Theorem \ref{theorem2}, we obtain that
any agent outside of the basic bicomponents converges to a convex combination
of the synchronized trajectories of the basic bicomponents. The
parameters of the convex combination only depend on the structure of the graph.


			\begin{figure}[pt]
				\includegraphics[width=9cm]{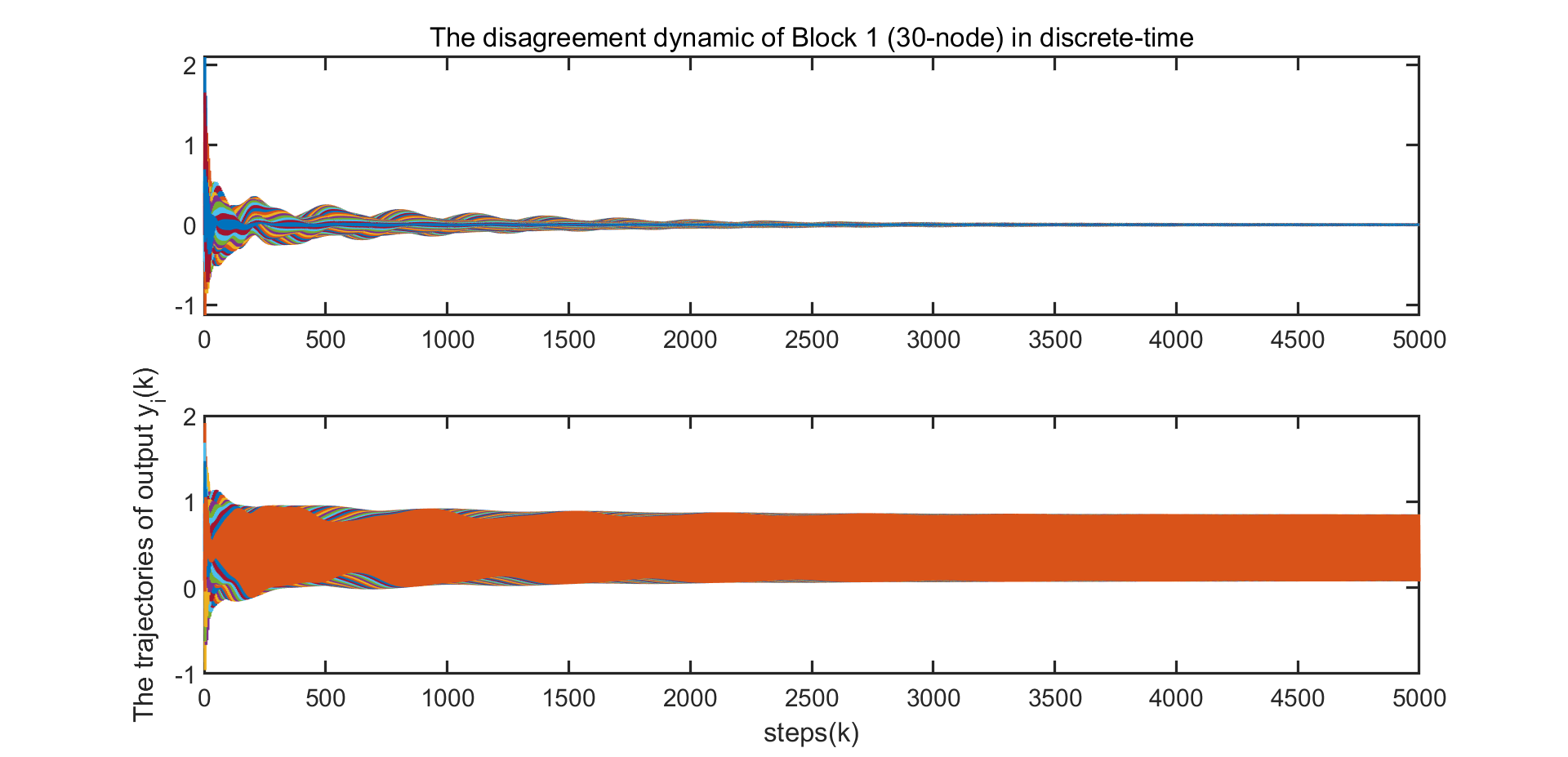}
				\centering
				\caption{30-node basic bicomponent for discrete-time MAS: disagreement dynamic among the agents and synchronized output trajectories.}\label{dt30cl}
			\end{figure}



\subsection{Discrete-time case}
Consider a discrete-time homogeneous MAS with agent model \eqref{eq1} give by:
\begin{equation*}
	A=\begin{pmatrix}
		0&1&1\\-1&0&1\\0&0&1
	\end{pmatrix},\ B=\begin{pmatrix}
		0\\0\\1
	\end{pmatrix},\ C=\begin{pmatrix}
		1&0&0
	\end{pmatrix}.
\end{equation*}
According to the design of \cite{liu-saberi-stoorvogel-tac-2023}, we obtain the following non-collaborative protocols for $i=1,\dots,N$,
			\begin{equation}\label{pro-lin-partial-ex-dis}
				\begin{system}{cl}
					{\chi}_i(k+1)&=\begin{pmatrix}
						-0.5&1&1\\-0.5&0&1\\-0.4&0&1
					\end{pmatrix}\chi_i(k)+\begin{pmatrix}
						0.5\\-0.5\\0.4
					\end{pmatrix}{\zeta}_i(k),\\
					u_i(k)&=-\begin{pmatrix}
						0&-0.1&0.1
					\end{pmatrix}\chi_i(k).
				\end{system}
			\end{equation}
			
			%
			%

			Then, we consider scale-free weak output synchronization result for the
			60-node discrete-time heterogeneous network shown in figure \ref{f4}.

\begin{figure}[t]
				\includegraphics[width=9cm]{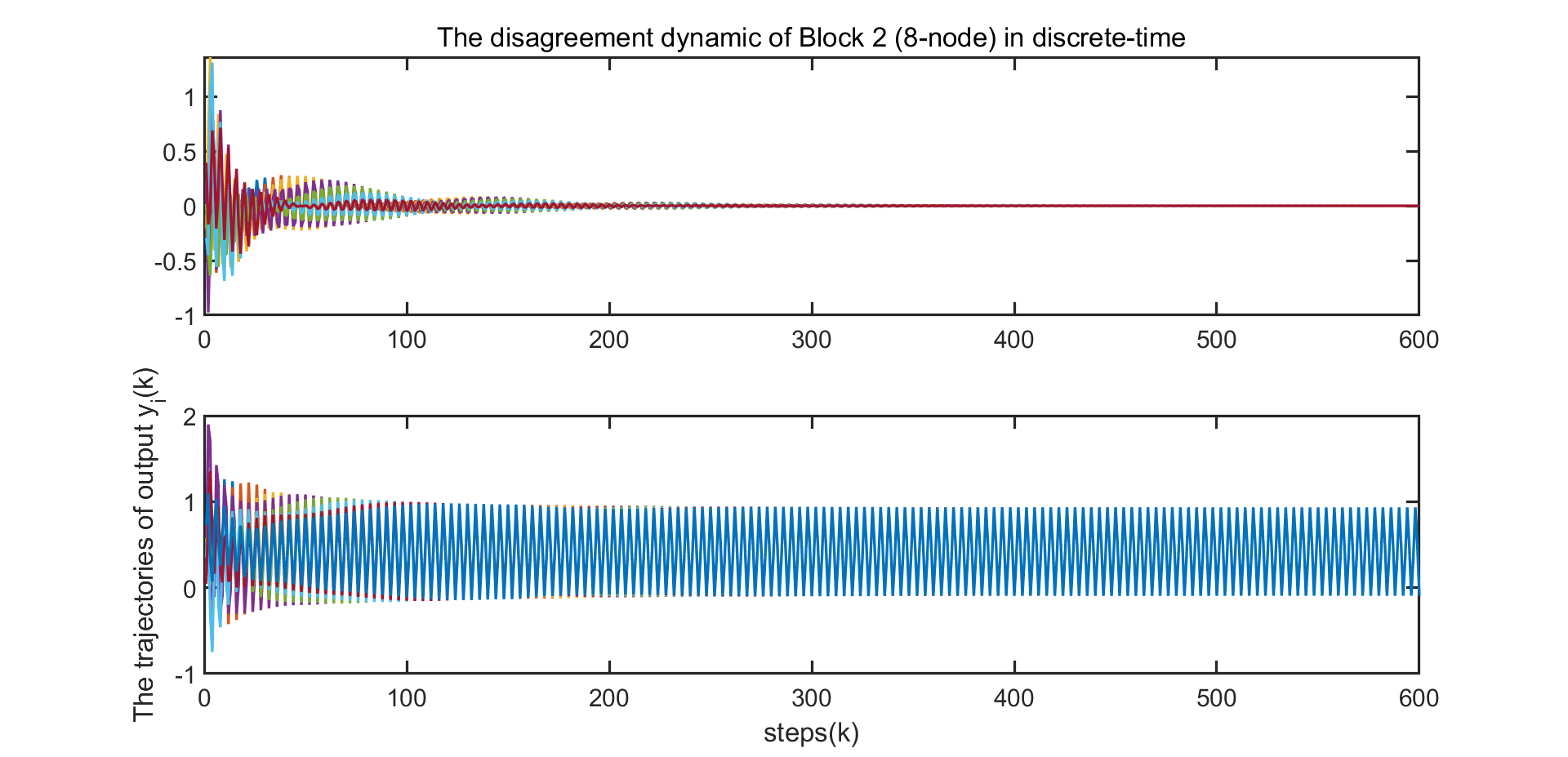} \centering
				\caption{8-node basic bicomponent for discrete-time MAS: disagreement dynamic among the agents and synchronized output trajectories.}\label{dt8cl}
			\end{figure}

\begin{figure}[t]
				\includegraphics[width=9cm]{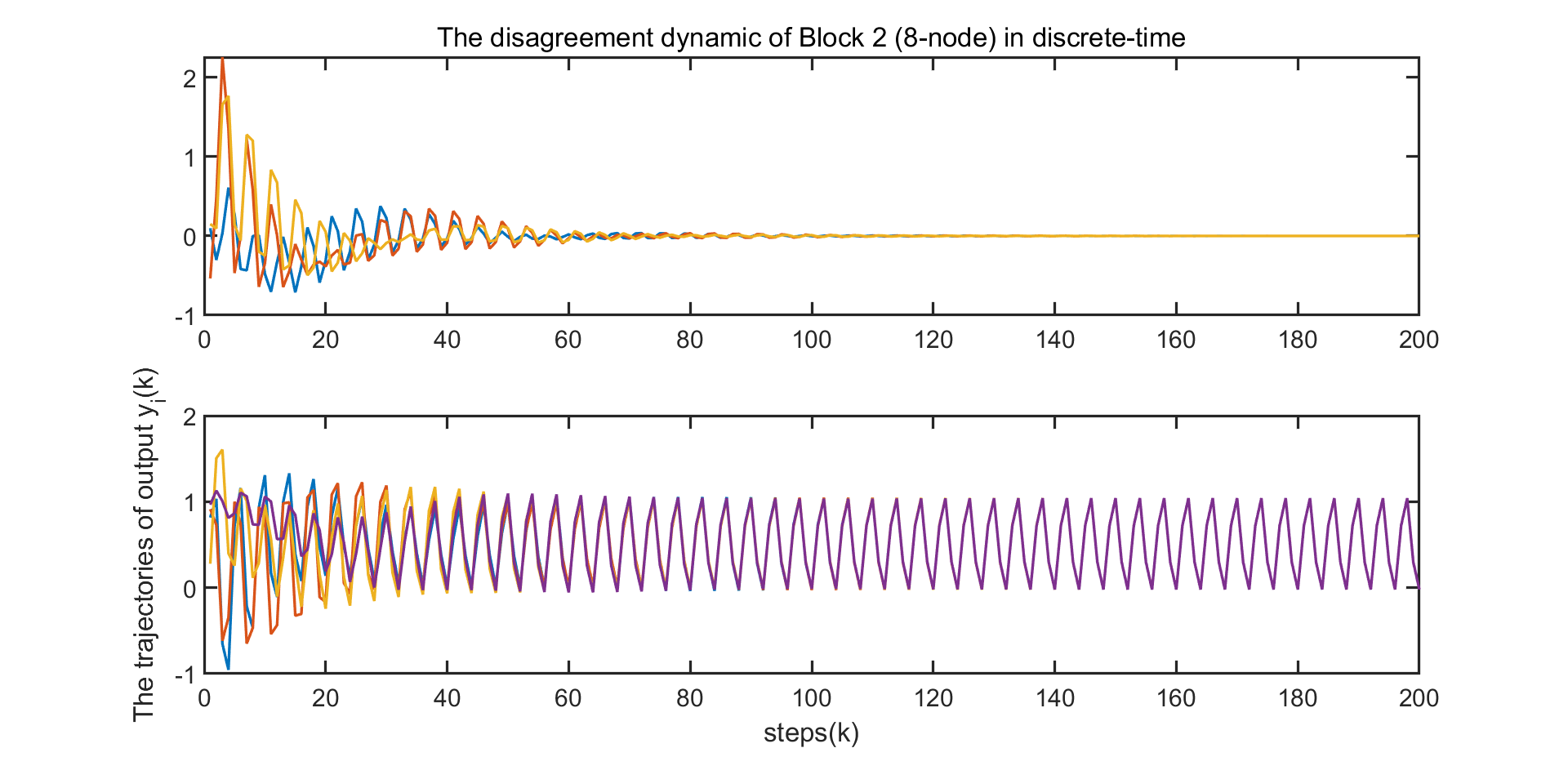}
				\centering
				\caption{4-node basic bicomponent for discrete-time MAS: disagreement dynamic among the agents and synchronized output trajectories.}\label{dt4cl}
			\end{figure}

By using the scale-free protocol \eqref{pro-lin-partial-ex-dis}, we
see that, consistent with the theory, we get output synchronization
the three basic bicomponents as illustrated in  figures \ref{dt30cl},
\ref{dt8cl} and \ref{dt4cl} respectively. Obviously, the disagreement
dynamic among the agents (the errors between the output of agents)
goes to zero within each basic bicomponent.

According to Theorem \ref{theorem2d}, we obtain that
any agent outside of the basic bicomponents converges to a convex combination
of the synchronized trajectories of the basic bicomponents. The
parameters of the convex combination only depend on the structure of
the graph.



\section{Conclusion}

In this paper we have shown how scale-free protocols which achieve
state synchronization behave when, due to a fault, the network no longer
contains a directed spanning tree. We have seen that the protocols
guarantee a stable response to these faults where agents converge to a
convex combination of the asymptotic behavior achieved in the basic
bicomponents. Moreover, this behavior is independent of the specific
scale-free protocols being used.

We think this form of synchronization is really useful in many
applications where network information is simply not available. For
instance we are currently working on extending these ideas in the
context of agents subject to disturbances.

\bibliographystyle{plain}
\bibliography{referenc}

\end{document}